\documentclass[12pt,a4paper,final]{revtex4}

\usepackage{sidecap}
\usepackage{ulem}
\usepackage{epsfig}
\usepackage{amsmath,amssymb,amsthm}
\usepackage{graphicx}
\usepackage{bm}
\usepackage{color}

\DeclareMathAlphabet{\mathpzc}{OT1}{pzc}{m}{it}

\begin{document}

\renewcommand{\textfraction}{0.00}


\newcommand{\vAi}{{\cal A}_{i_1\cdots i_n}}
\newcommand{\vAim}{{\cal A}_{i_1\cdots i_{n-1}}}
\newcommand{\vAbi}{\bar{\cal A}^{i_1\cdots i_n}}
\newcommand{\vAbim}{\bar{\cal A}^{i_1\cdots i_{n-1}}}
\newcommand{\htS}{\hat{S}}
\newcommand{\htR}{\hat{R}}
\newcommand{\htB}{\hat{B}}
\newcommand{\htD}{\hat{D}}
\newcommand{\htV}{\hat{V}}
\newcommand{\cT}{{\cal T}}
\newcommand{\cM}{{\cal M}}
\newcommand{\cMs}{{\cal M}^*}
\newcommand{\vk}{\vec{\mathbf{k}}}
\newcommand{\bk}{\bm{k}}
\newcommand{\kt}{\bm{k}_\perp}
\newcommand{\kp}{k_\perp}
\newcommand{\km}{k_\mathrm{max}}
\newcommand{\vl}{\vec{\mathbf{l}}}
\newcommand{\bl}{\bm{l}}
\newcommand{\bK}{\bm{K}}
\newcommand{\bb}{\bm{b}}
\newcommand{\qm}{q_\mathrm{max}}
\newcommand{\vp}{\vec{\mathbf{p}}}
\newcommand{\bp}{\bm{p}}
\newcommand{\vq}{\vec{\mathbf{q}}}
\newcommand{\bq}{\bm{q}}
\newcommand{\qt}{\bm{q}_\perp}
\newcommand{\qp}{q_\perp}
\newcommand{\bQ}{\bm{Q}}
\newcommand{\vx}{\vec{\mathbf{x}}}
\newcommand{\bx}{\bm{x}}
\newcommand{\tr}{{{\rm Tr\,}}}
\newcommand{\bc}{\textcolor{blue}}

\newcommand{\beq}{\begin{equation}}
\newcommand{\eeq}[1]{\label{#1} \end{equation}}
\newcommand{\ee}{\end{equation}}
\newcommand{\bea}{\begin{eqnarray}}
\newcommand{\eea}{\end{eqnarray}}
\newcommand{\beqar}{\begin{eqnarray}}
\newcommand{\eeqar}[1]{\label{#1}\end{eqnarray}}

\newcommand{\half}{{\textstyle\frac{1}{2}}}
\newcommand{\ben}{\begin{enumerate}}
\newcommand{\een}{\end{enumerate}}
\newcommand{\bit}{\begin{itemize}}
\newcommand{\eit}{\end{itemize}}
\newcommand{\ec}{\end{center}}
\newcommand{\bra}[1]{\langle {#1}|}
\newcommand{\ket}[1]{|{#1}\rangle}
\newcommand{\norm}[2]{\langle{#1}|{#2}\rangle}
\newcommand{\brac}[3]{\langle{#1}|{#2}|{#3}\rangle}
\newcommand{\hilb}{{\cal H}}
\newcommand{\pleft}{\stackrel{\leftarrow}{\partial}}
\newcommand{\pright}{\stackrel{\rightarrow}{\partial}}

\title{Complex suppression patterns distinguish between major energy loss effects in Quark-Gluon Plasma}

\author{Magdalena Djordjevic}
\affiliation{Institute of Physics Belgrade, University of Belgrade, Serbia}

\begin{abstract} Interactions of high momentum partons with Quark-Gluon Plasma created in relativistic heavy-ion collisions provide an excellent tomography tool for this new form of matter. Recent measurements for charged hadrons and unidentified jets at the LHC show an unexpected flattening of the suppression curves at high momentum, exhibited when either momentum or the collision centrality is changed. Furthermore, a limited data available for B probes indicate a qualitatively different pattern, as nearly the same flattening is exhibited for the curves corresponding to two opposite momentum ranges. We here show that the experimentally measured suppression curves are well reproduced by our theoretical predictions, and that the complex suppression patterns are due to an interplay of collisional, radiative energy loss and the dead-cone effect. Furthermore, for B mesons, we predict that the uniform flattening of the suppression indicated by the limited dataset is in fact valid across the entire span of the momentum ranges, which will be tested by the upcoming experiments. Overall, the study presented here, provides a rare opportunity for pQCD theory to qualitatively distinguish between the major energy loss mechanisms at the same (nonintuitive) dataset.
\end{abstract}


\maketitle

\section{Introduction}

\begin{figure*}
\epsfig{file=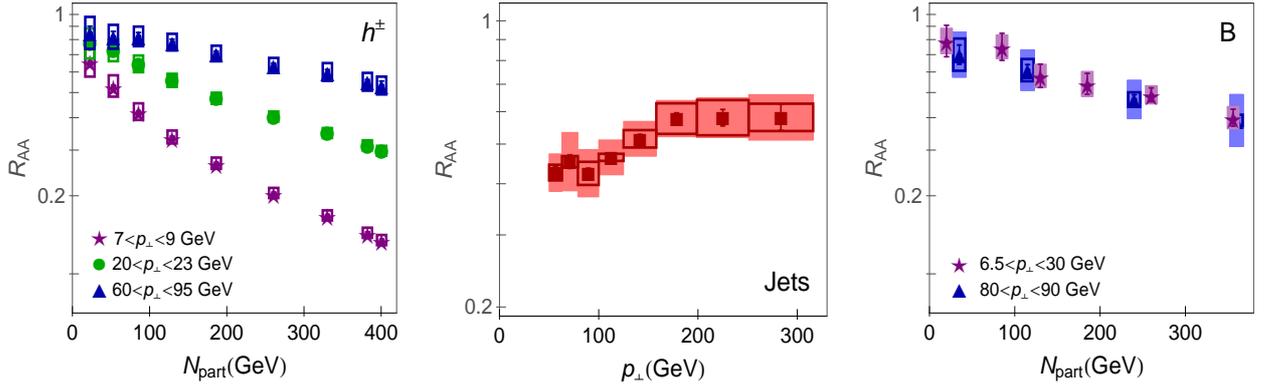,width=6.5in,height=2.15in,clip=5,angle=0}
\vspace*{-0.4cm}
\caption{{\bf Suppression patterns at the LHC.} {\it Left panel:} ATLAS experimental data for $h^\pm$~\cite{ATLAS_CH} $R_{AA}$ {\it vs.} $N_{part}$ are shown, where purple stars, green dots and blue triangles correspond, respectively, to the data for $7<p_\perp<9$, $20<p_\perp<23$ and $60<p_\perp<95$ GeV momentum regions. {\it Central panel:} ATLAS~\cite{ATLAS_Jets} most central experimental data for unidentified jet $R_{AA}$ {\it vs.} $p_\perp$ are shown (red squares). {\it Right panel:} $R_{AA}$ {\it vs.} $N_{part}$ CMS experimental data are shown for non-prompt $J/\Psi$ (purple stars with $6.5<p_\perp<30$ GeV)~\cite{CMS_JPsi} and B jets (blue triangles with $80<p_\perp<90$ GeV)~\cite{CMS_Bjets}.}
\label{Data}
\end{figure*}

In the collisions of ultra-relativistic heavy ions at RHIC and LHC experiments, a new state of matter, called Quark-Gluon Plasma (QGP), is created. Rare high momentum probes transverse and interact with the medium, providing an excellent QGP tomography tool~\cite{Bjorken}. Utilizing such tool requires comparing experimental data with theoretical predictions, where nonintuitive observations present a particular challenge for the theory. Such a challenge is provided by the recent measurements of suppression for charged hadrons~\cite{ATLAS_CH}, unidentified jets~\cite{ATLAS_Jets} and B probes~\cite{CMS_JPsi,CMS_Bjets} at 2.76 TeV Pb+Pb collisions at the LHC. In particular, in Fig.~\ref{Data} (the left and the central panels) are shown ATLAS~\cite{ATLAS_CH,ATLAS_Jets} suppression ($R_{AA}$) data for different momentum ranges and as a function of both the number of participants ($N_{part}$), see the left panel, and momentum ($p_\perp$), see the central panel. In particular, ATLAS charged hadron ($h^\pm$) data~\cite{ATLAS_CH} show that $R_{AA}$ {\it vs.} $N_{part}$ curves become increasingly flatter as one moves towards higher momentum ranges (compare purple, green and blue data points in the left panel). Furthermore, the central panel shows flattening (saturation) of $R_{AA}$ at high momentum, that can be observed for $R_{AA}$ {\it vs.} $p_\perp$ dependence corresponding to unidentified jets at ATLAS (red squares)~\cite{ATLAS_Jets}. These observations are highly non-trivial: the left panel suggest that, while the lower momentum light flavor probes are very sensitive to the $N_{part}$ - and consequently to the system size and energy density- such sensitivity is significantly smaller for the high momentum probes. For the central panel, one observes an apparent plateau reached by $R_{AA}$ data at high $p_\perp$, leading to the question what energy loss mechanism is responsible for this effect. Moreover, a qualitatively different $R_{AA}$ {\it vs.} $N_{part}$ pattern is apparently observed for B mesons: $R_{AA}$ for non-prompt $J/\Psi$ at lower momentum~\cite{CMS_JPsi} and B jets~\cite{CMS_Bjets} at high momentum (the purple and the blue dots in the right panel of Fig.~\ref{Data}, respectively), surprisingly show the same $R_{AA}$ {\it vs.} $N_{part}$ for these two opposite momentum ranges  - both of them indicating small sensitivity to the increase in $N_{part}$; for observing the difference with $h^\pm$ data, compare the purple and the blue data points on the left and the right panels of Fig.~\ref{Data}.

We will study the data patterns summarized above within our state-of-the-art dynamical energy loss formalism~\cite{MD_PRC,DH_PRL}. Briefly, the formalism takes into account that QGP consists of dynamical (moving) partons - which removes the widely used assumption of static scattering centers - and that the created medium has a finite size. Both collisional~\cite{MD_Coll} and radiative~\cite{MD_PRC,DH_PRL} energy losses are calculated within the same theoretical framework, which is applicable to both light and heavy flavor, and includes finite magnetic mass~\cite{MD_MagnMass} and running coupling~\cite{MD_PLB}. This formalism is integrated in a numerical procedure which takes into account the up-to-date initial distributions~\cite{Cacciari:2012,Vitev0912}, fragmentation functions~\cite{DSS}, path-length~\cite{WHDG,Dainese} and multi-gluon fluctuations~\cite{GLV_MG}; importantly {\it no free parameters} are used in comparing the model predictions with the data. Our aim is not only showing that the dynamical energy loss formalism can well explain the complex $R_{AA}$ data patterns, but even more providing an intuitive explanation for the unexpected experimental observations, which may point to an anticipated example of a qualitative interplay between the major energy loss effects. Finally, we will provide predictions for the upcoming experimental measurements that will further test the mechanism proposed here. Note that predictions presented here are applicable to both 2.76 TeV and 5.02 TeV Pb+Pb collisions at the LHC, since in~\cite{DBZ, MD_5TeV}, we predict that $R_{AA}$s at these two collision energies will be the same.

\section{Results}

\begin{figure*}
\epsfig{file=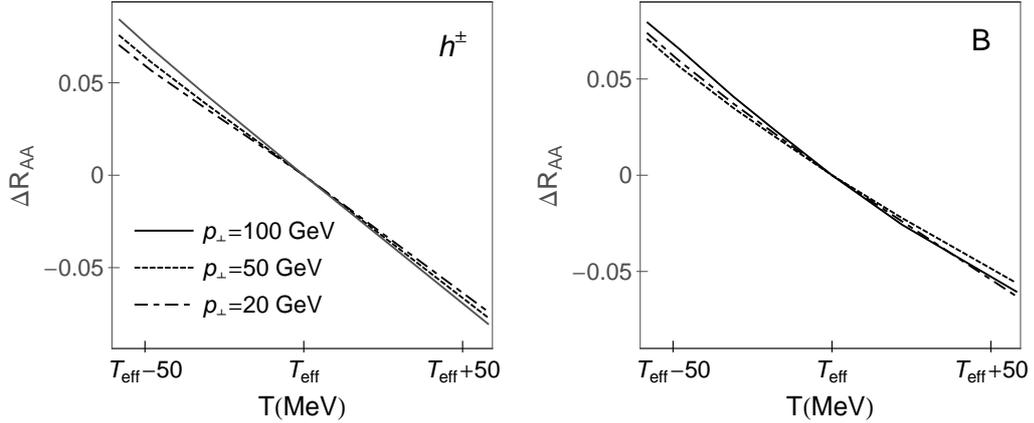,width=5.3in,height=2.55in,clip=5,angle=0}
\vspace*{-0.4cm}
\caption{{\bf $R_{AA}$ dependence on the average temperature of QGP.} {\it Left panel:} The difference in $R_{AA}$ at temperature $T$ and at effective temperature $T_{eff}$ ($\Delta R_{AA} = R_{AA}(T) -R_{AA}(T_{eff})$) as a function of QGP temperature is shown for charged hadrons and B mesons on the left and right panels, respectively. On each panel full, dashed and dot-dashed curves, respectively, correspond to $p_\perp$ of 100 GeV, 50 GeV and 20 GeV. Magnetic to electric mass ratio is set to $\mu_M/\mu_E =0.4$, while $T_{eff}=304$~MeV corresponds to the face value of the effective temperature extracted from ALICE~\cite{ALICE_T}. }
\label{TDepLB}
\end{figure*}

The predictions are generated by the dynamical energy loss formalism, where the computational procedure and the parameter set are described in detail in~\cite{MD_PLB}. Briefly, we consider a QGP  with $n_f{\,=\,}3$ and $\Lambda_{QCD}=0.2$~GeV. For the light quarks, we assume that their mass is dominated by the thermal mass $M{\,=\,}\mu_E/\sqrt{6}$, where the temperature dependent Debye mass $\mu_E (T)$  is obtained from~\cite{Peshier}, while the gluon mass is  $m_g=\mu_E/\sqrt{2}$~\cite{DG_TM} and the charm (bottom) mass is $M{\,=\,}1.2$\,GeV ($M{\,=\,}4.75$\,GeV). Since various non-perturbative calculations~\cite{Maezawa,Nakamura,Hart,Bak} have shown that magnetic mass $\mu_M$ is different from zero in QCD matter created at the LHC and RHIC, the finite magnetic mass effect is also included in our framework. Moreover, from these non-perturbative QCD calculations it is extracted that magnetic to electric mass ratio is $0.4 < \mu_M/\mu_E < 0.6$, so the uncertainty in the predictions, presented in this section, will come from this range of screening masses ratio. Path-length distributions are taken from~\cite{Dainese}.

The temperatures for different centralities are calculated according to~\cite{DDB_PLB}. As a starting point in this calculation we use the effective temperature ($T_{eff}$) of 304 MeV for 0-40$\%$ centrality Pb+Pb collisions at the LHC~\cite{ALICE_T} experiments (as extracted by ALICE). As this temperature comes with $\pm 60$ MeV errorbar, we first ask how this uncertainty affects the calculated suppression. Conseqeuntly, in Fig.~\ref{TDepLB} we show how the variations (uncertainty in the average QGP temperature) influences the suppression results for different types of flavor (both light and heavy), and at different $p_\perp$ regions. We see that $R_{AA}$ dependence on the average temperature of QGP is almost linear. We also see that the change in the average temperature of the QGP does not significantly affect the suppression, i.e. maximal temperature uncertainly (of 60 MeV), leads to the change in $R_{AA}$ of less than 0.07. Furthermore, we see that dependence of $R_{AA}$ on $T_{eff}$ is almost the same for all parton energies and all types of flavor. We therefore conclude that this uncertainty in the effective temperature would basically lead to a systematic (constant value) shift in the predictions, so the results presented in this paper would not be affected by this uncertainty. Furthermore, extensive comparison~\cite{MD_PLB,MD_PRL,DDB_PLB} of our theoretical predictions with experimental data (corresponding to different probes, experiments and centrality regions), shows a robust agreement when the experimentally measured average QGP temperature of $T_{eff}=304$~MeV is used, so we will further use this temperature as a starting point in the prediction calculations. Furthermore, note that these extensive comparisons use the same theoretical framework and the parameter set (corresponding to the standard literature values) as the predictions presented in this paper; consequently, the predictions presented here are well constrained, not only by the absence of the free parameters, but also by the agreement with an extensive set of other data.

\begin{figure*}
\epsfig{file=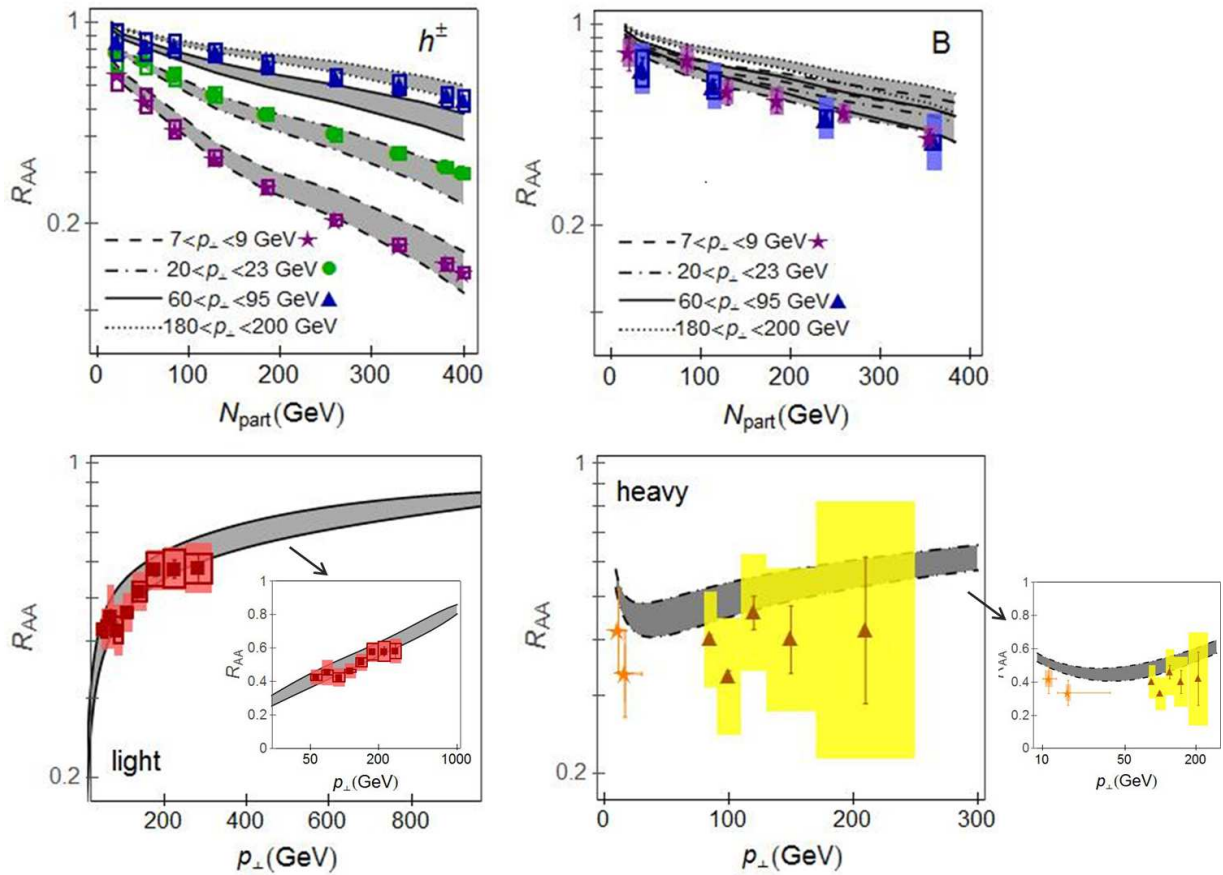,width=6.3in,height=4.7in,clip=5,angle=0}
\vspace*{-0.4cm}
\caption{{\bf Suppression patterns at the LHC: comparison of theoretical predictions with experimental data} {\it Upper left panel:} Theoretical predictions for $R_{AA}$ {\it vs.} $N_{part}$ are compared with ATLAS experimental data for $h^\pm$~\cite{ATLAS_CH}, where purple stars, green dots and blue triangles correspond, respectively, to the data for $7<p_\perp<9$, $20<p_\perp<23$ and $60<p_\perp<95$ GeV momentum regions. Gray bands with dashed, dot-dashed, full and dotted boundaries correspond, respectively, to the predictions for $7<p_\perp<9$, $20<p_\perp<23$, $60<p_\perp<95$, $180<p_\perp<200$ GeV momentum regions.  {\it Upper right panel:} Theoretical predictions for $R_{AA}$ {\it vs.} $N_{part}$ are compared with CMS experimental data for non-prompt $J/\Psi$ (purple stars with $6.5<p_\perp<30$ GeV)~\cite{CMS_JPsi} and B jets (blue triangles with $80<p_\perp<90$ GeV)~\cite{CMS_Bjets}. The gray bands are equivalent to those in the left panel.
{\it Lower left panel:} Theoretical predictions for $R_{AA}$ {\it vs.} $p_\perp$ are compared with ATLAS (red squares)~\cite{ATLAS_Jets} most central experimental data for unidentified jets. Insert corresponds to the same figure plot on the logarithmic scale. {\it Lower right panel:} Theoretical predictions for $R_{AA}$ {\it vs.} $p_\perp$ are compared with CMS experimental data for non-prompt $J/\Psi$~\cite{CMS_JPsi} (orange stars) and b-jets~\cite{CMS_Bjets} (brown triangles). Insert corresponds to the same figure plot on the logarithmic scale. On each panel, the upper (lower) boundary of each gray band corresponds to $\mu_M/\mu_E =0.4$ ($\mu_M/\mu_E =0.6$). }
\label{DataVsTheory}
\end{figure*}
In the upper left panel of Fig.~\ref{DataVsTheory}, we provide predictions which agree very well with the ATLAS $h^\pm$ data~\cite{ATLAS_CH} for three different momentum ranges ($7<p_\perp<9$, $20<p_\perp<23$, $65<p_\perp<90$~GeV). The predicted curves reproduce well the tendency observed in the data, i.e. as one moves to higher energy ranges, $R_{AA}$ {\it vs.} $N_{part}$ becomes increasingly flatter. Another tendency of $R_{AA}$ {\it vs.} $N_{part}$ is also apparent from the predictions, i.e. as one moves towards higher momentum ranges, the difference between the curves becomes increasingly smaller; we will call this the apparent saturation in $R_{AA}$ {\it vs.} $N_{part}$ curves.

In the lower left panel of Fig.~\ref{DataVsTheory}, we see that our predictions agree well with the measured~\cite{ATLAS_Jets} $R_{AA}$ {\it vs.} $p_\perp$ dependance. Moreover, we see that the "plateau"~\cite{ATLAS_Jets}, often referred as surprising, corresponds to the slow increase in the predicted curves, which we further call saturation in $R_{AA}$ {\it vs.} $p_{\perp}$ dependence. From the insert in this panel (where we use the logarithmic scale for $p_\perp$ and linear scale for $R_{AA}$), we see that this slow increase in $R_{AA}$ corresponds to the linear dependence on $\ln (p_\perp)$, as can be observed from both the experimental data and the theoretical predictions. Finally, in the upper right and the lower right panels of Fig.~\ref{DataVsTheory}, we see that our predictions can also well reproduce the experimental data~\cite{CMS_JPsi,CMS_Bjets} for B probes, which indicate qualitatively substantially different pattern compared to $h^\pm$ data. Moreover, the calculated $R_{AA}$ {\it vs.} $N_{part}$ and $R_{AA}$ {\it vs.} $p_\perp$ are largely flat across the {\it entire} span of the momentum ranges, and the apparent saturation in $R_{AA}$ {\it vs.} $N_{part}$ curves - which is for light probes observed only at higher $p_\perp$ range - is for B probes predicted for the entire momentum span. As a digression, we here note that, while our predictions are generated for single particles, the corresponding experimental measurements for high $p_\perp$ single particles are not always available. Consequently, we here compare our predictions with both single particles and (in some cases) with jets. While single particles and jets are not equivalent observables, we think that such comparison is not unreasonable because both theoretical predictions and experimental data (when available) indicate an overlap (within errorbars) between single particle and jet data~\cite{DBZ}. This gives us confidence that, when high $p_\perp$ single particle data become available at 5 TeV collision energy, these experimental data will likely largely overlap with the existing jet $R_{AA}$ data.

The predictions shown in Fig.~\ref{DataVsTheory} contain several features, similar to those indicated by the experimental data (see the Introduction). First, the flattening and the apparent saturation observed in $R_{AA}$ {\it vs.} $N_{part}$ curves for $h^\pm$ imply that, at high $p_\perp$, the predictions indicate significantly smaller sensitivity to the collision centrality, and consequently to the corresponding changes in the medium properties; this is in contrast to the lower $p_\perp$, where predictions exhibit a considerable sensitivity to the collision centrality. Furthermore, the saturation observed in $R_{AA}$ {\it vs.} $p_\perp$ predictions - consistent with the corresponding plateau in the experimental data - indicates an unexpectedly slow change of jet energy loss with the initial jet energy at high $p_\perp$. Finally, the $R_{AA}$ {\it vs.} $N_{part}$ pattern predicted for the B probes is also surprising: here, a qualitatively different pattern compared to the light probes is obtained, where B probes show small sensitivity to the medium properties across the entire momentum span - as opposed to the small $h^\pm$ sensitivity at only high $p_\perp$. As the available data for B probes indicated in the right panels are limited (and indirect, i.e. corresponding to the different observables), note that the calculated B meson results also correspond to novel predictions (expected to become available at the 5 TeV collision energies), whose comparison with the upcoming data will test how our formalism can explain qualitatively unexpected observations.

\begin{figure*}
\epsfig{file=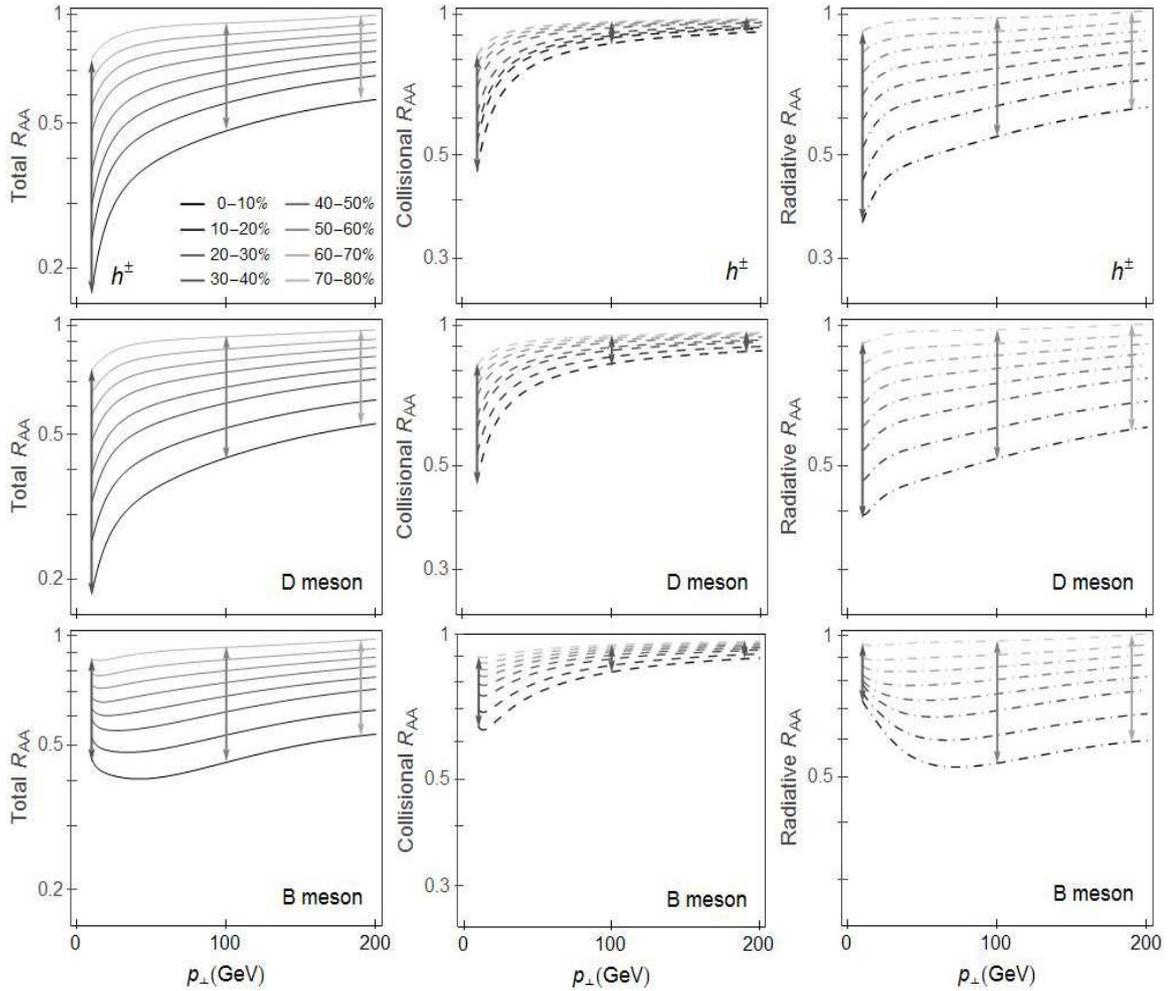,width=6.in,height=5.17in,clip=5,angle=0}
\vspace*{-0.4cm}
\caption{{\bf $R_{AA}$ {\it vs.} $p_\perp$ for different centrality regions.} Left, central and right panels show, respectively, total, collisional and radiative contributions to $R_{AA}$. Upper, middle and lower panels correspond to $h^\pm$, D and B mesons, respectively. On each panel, different curves correspond to different centrality regions, as denoted by the legend in the upper left panel ($\mu_M/\mu_E$ is set to 0.4). Also, on each panel, three double-sided arrows represent the $R_{AA}$ spans for, respectively, $p_\perp$ of 10, 100 and 190~GeV. }
\label{LightBottomPatterns}
\end{figure*}

We therefore aim understanding the nonintuitive patterns in the predictions/data outlined above. For the light probes, this explanation is provided by the upper left panel of Fig.~\ref{LightBottomPatterns}, which shows $R_{AA}$ {\it vs.} $p_{\perp}$ dependence for a family of curves corresponding to increasing collision centrality. Note two main properties of these curves: First, their shape, which leads to the saturation in $R_{AA}$ {\it vs.} $p_{\perp}$ exhibited in the central panel in Fig.~\ref{DataVsTheory}; this shape will be explained by the upper central and the right panels in Fig.~\ref{LightBottomPatterns}. Secondly, their density, which (non-uniformly) increases as one moves from lower to high $p_\perp$ - with that respect, it may be useful to observe $R_{AA}$ {\it vs.} $p_\perp$ curves as field flux lines. For visualizing how the relevant curve density changes, three vertical arrows are indicated in the upper left panel of Fig.~\ref{LightBottomPatterns} - these arrows relate to understanding $h^\pm$ $R_{AA}$ {\it vs.} $N_{part}$ predictions in the upper left panel of Fig.~\ref{DataVsTheory}. Specifically, the leftmost arrow, corresponding to lower ($\sim 10$ GeV) $p_\perp$, spans a much larger $R_{AA}$ range compared to the two right arrows, which correspond to higher $p_\perp$. This observation directly translates to the fact that $R_{AA}$ {\it vs.} $N_{part}$ curves are much steeper at lower, compared to high, $p_\perp$ ranges. Moreover, there is a much larger difference in $R_{AA}$ span between the leftmost and the central arrows, as compared to the central and the rightmost arrows; this being despite the fact that the three arrows are spaced equidistantly in momentum. A direct consequence of these differences in $R_{AA}$ span, is the apparent saturation in $R_{AA}$ {\it vs.} $N_{part}$ in the upper left panel of Fig.~\ref{DataVsTheory}, i.e. the fact that there is an increasingly smaller difference between $R_{AA}$ {\it vs.} $N_{part}$ curves as one moves towards increasingly higher momentum ranges.

The shape of the total $R_{AA}$ {\it vs.} $p_\perp$ curves (leading to the saturation in the lower left panel of Fig.~\ref{DataVsTheory}) is a consequence of an interplay between the collisional and the radiative contributions to the suppression~\footnote{The total $R_{AA}$ is approximately (though not exactly) equal to the product of the radiative and collisional contributions}. As can be seen at the upper central panel of Fig.~\ref{LightBottomPatterns}, the collisional contribution to the $R_{AA}$ is notable for smaller $p_\perp$, where it increases steeply with momentum, rapidly approaching 1 at high $p_\perp$, providing a small contribution to total $R_{AA}$ at high $p_\perp$ region. On the other hand, the radiative contribution decreases much slower with the momentum, and has a significant contribution to $R_{AA}$ even at higher $p_\perp$. Consequently, the steep increase in total $R_{AA}$ at lower $p_\perp$ is driven by the dominant collisional contribution to $R_{AA}$ in that momentum range, while the slow increase (apparent saturation) of $R_{AA}$ at high $p_\perp$ is due to the dominant radiative contribution. This interplay then explains the saturation (plateau) of $R_{AA}$ observed at high $p_\perp$. Moreover, such interplay between the collisional and the radiative contributions to $R_{AA}$, also determines the density of $R_{AA}$ {\it vs.} $p_\perp$ curves. As can be seen from the vertical arrows indicated in the upper central and the right panels in Fig.~\ref{LightBottomPatterns}, the collisional contribution is responsible for the large span of total $R_{AA}$ with changing centrality at lower momentum. On the other hand, at higher momentum, the radiative contribution exhibits a significantly smaller and largely uniform $R_{AA}$ span, therefore resulting in the larger and more uniform total $R_{AA}$ curve density in that momentum range.

In the central row of Figure~\ref{LightBottomPatterns}, we see that D mesons show the similar behavior as $h^\pm$. Therefore, for the purpose of analyzing different suppression patterns at the LHC, D mesons can be used as an alternative to $h^\pm$. While D mesons are experimentally harder to measure, from theoretical perspective they have a clear advantage over $h^\pm$. This is because $h^\pm$s are composed of both light quarks and gluons, so that $h^\pm$ presents an indirect probe of light flavor, which is significantly influenced by the fragmentation functions. On the other hand, D meson $R_{AA}$ is a clear probe of bare charm quark $R_{AA}$, i.e. D meson $R_{AA}$ is not influenced by fragmentation functions. Therefore, despite the obvious experimental difficulty in measuring the D meson (compared to $h^\pm$) $R_{AA}$ patterns, D meson has an obvious advantage over $h^\pm$ for analyzing the unintuitive interplays of collisional {\it vs.} radiative energy loss and dead-cone effect~\cite{Kharzeev}, discussed in this paper.

An intuitive explanation behind the different $R_{AA}$ {\it vs.} $N_{part}$ pattern observed for B probes (the right panel in Fig. 1) is provided by the lower panels of Fig.~\ref{LightBottomPatterns}. In distinction to $h^\pm$, in the lower left panel, we see that total $R_{AA}$ {\it vs.} $p_\perp$ curves have mostly uniform density, with a largely flat shape of the curves across the entire momentum range. The difference with respect to $h^\pm$ is clearly due to the radiative contribution to the total $R_{AA}$, as the curves corresponding to the B meson collisional contribution are largely equivalent to those for $h^\pm$ (compare the upper and lower central panels in Fig.\ref{LightBottomPatterns}). In particular, note an unusual shape of radiative $R_{AA}$ {\it vs.} $p_\perp$ curves (the lower right panel), which is a consequence of a strong dead-cone effect~\cite{Kharzeev} in bottom quark energy loss. This unusual shape leads to a large density of radiative $R_{AA}$ {\it vs.} $p_\perp$ curves at lower momentum, and to a largely uniform curve density for high momentum. As a consequence, for lower momentum, the effect of the relatively large $R_{AA}$ span for the collisional contribution is abolished by the small $R_{AA}$ span for the radiative contribution, this leading to the flat and uniform density for total $R_{AA}$ {\it vs.} $p_\perp$ curves observed in the lower left panel of Fig.\ref{LightBottomPatterns}. Such curve shape then evidently leads to a largely uniform total $R_{AA}$ span across the whole range of momentum, as indicated by the three vertical arrows in the lower left panel of Fig.\ref{LightBottomPatterns}, which then leads to our predictions of the largely flat $R_{AA}$ {\it vs.} $p_\perp$ and almost overlapping $R_{AA}$ {\it vs.} $N_{part}$ curves for B mesons. Comparison of theoretical predictions with experimental data shown in the lower (left and right) panels of Fig.~\ref{DataVsTheory} show an indication that these predictions might be in accordance with experimental data. However, the data shown in Fig.~\ref{DataVsTheory} are indirect, very limited and correspond to different bottom observables, so more detailed experimental data at 5 TeV Pb+Pb collisions at the LHC are needed to confirm (or dispute) the predictions presented in this study.

\section{Conclusion} A starting point for this work is an observation of a plateau reached at high momentum for $R_{AA}$ {\it vs.} $p_\perp$ measurements. Starting from this observation, we here combined related experimental data, which reveal an unexpected pattern in the $R_{AA}$ data. We showed that these data patterns are well reproduced by the theoretical predictions for charged hadrons and unidentified jets, with no free parameters used. For B mesons, we predict that the tendency indicated by the limited available data will be exhibited across the entire span of the momentum ranges - this  prediction will be tested by the upcoming experimental data expected from the 5 TeV Pb+Pb collisions at the LHC.

We showed that these complex data patterns have, in fact, a simple qualitative interpretation, where it is useful to observe $R_{AA}$ {\it vs.} $p_\perp$ curves as field flux lines whose density changes across different momentum. These curve properties - which lead to the unexpected dependence of $R_{AA}$ on $N_{part}$ and $p_\perp$, and to qualitatively different $R_{AA}$ patterns for the light and heavy (i.e. bottom) probes - are determined by an interplay of collisional, radiative energy loss and the dead-cone effect.

Consequently, the results presented here provide a rare opportunity to qualitatively assess how the theory can account for two crucial effects: First, different suppression patterns exhibited by different probe types (here B mesons vs. charged hadrons or D mesons), providing a clear test of the dead cone effect. Second, contributions of different energy loss mechanisms, providing a test of an interplay between the collisional and the radiative energy loss. This point is even more important having in mind extensive experimental efforts aimed at assessing contributions of various energy loss effects. Consequently, this study provides both an important test of explaining and predicting complex data patterns and a clear qualitative example for distinguishing between major energy loss mechanisms.

{\em Acknowledgments:}
This work is supported by Marie Curie IRG within the $7^{th}$ EC Framework Programme (PIRG08-GA-2010-276913) and by the Ministry of Science of the Republic of Serbia
under project numbers ON173052 and ON171004. I thank B. Blagojevic for help with numerics and Marko Djordjevic for useful discussions.

\end{document}